\documentclass{aa}  

\usepackage{graphicx}
\usepackage{placeins}
\usepackage{txfonts}
\usepackage{ulem}
\usepackage{pdflscape}
\usepackage{amsmath}
\usepackage[caption=false]{subfig}

\usepackage{mathabx}
\usepackage{hyperref}
\usepackage{academicons}
\usepackage{xcolor}

\hypersetup{colorlinks,allcolors=blue}

\newcommand{\MEarth}{M$_{\mathrm{\oplus}}$\,}
\newcommand{\MJup}{M$_{\mathrm{Jup}}$\,}
\newcommand{\RJup}{R$_{\mathrm{Jup}}$\,}
\newcommand{\RSun}{R$_{\odot}$\,}
\newcommand{\MSun}{M$_{\odot}$\,}
\newcommand{\LSun}{L$_{\odot}$\,}
\newcommand{\teff}{$T_{\rm eff}$\,}

\newcommand{\kms}{km\, s$^{-1}$\,}

\begin{document} 

   \title{Analysis of the young disk around WRAY 15-1880: does it contain a primitive planetary system? }

   \author{Elisabetta Rigliaco\inst{1}, 
    Raffaele Gratton\inst{1}, 
    Silvano Desidera\inst{1}, 
    Gabriele Columba \inst{2},
    Enrico Grippi \inst{3}
          }
   \institute{INAF – Osservatorio Astronomico di Padova, Vicolo dell’Osservatorio 5, 35122 Padova, Italy \thanks{\email{elisabetta.rigliaco@inaf.it}}
    \and
    Dipartimento di Fisica e Astronomia “Augusto Righi”, Università di Bologna, Via Piero Gobetti 93/2, I-40129 Bologna, Italy
    \and
    Dipartimento di Fisica e Astronomia "Galileo Galilei", Universit\`a degli Studi di Padova, Vicolo dell’Osservatorio 3, 35122, Padova, Italy}

   \date{Received 27 January 2026; accepted 8 June 2026 }

  \abstract{Observations of (giant) planets accreting material within their natal environment are crucial to constrain models for their formation. WRAY 15-1880 (aka RX J1842.9-3532) in the Corona Australis (CrA) complex has a prominent pre-transitional disk, and an age of $2.8\pm 0.7$~Myr, computed by comparison with isochrones using the accurate dynamical mass derived from disk kinematics. Hence, this star is in the late phases of disk evolution and might host accreting planets. We acquire new polarimetric imaging data with VLT-SPHERE and analyze archive observations taken with VLT-SPHERE, VLT-MUSE, and ALMA, finding a candidate Jupiter-like companion within the disk gap from high-contrast imaging. The mass estimates of the candidate companion, derived from various methods, are consistent with an object in the range of 0.3--7.6~\MJup. The spectrum of the candidate companion is consistent with a T3 spectral type, in agreement with expectations of an object of a few jupiter masses. We find an emission blob North-West of the star in the ALMA data rotating solidly with the candidate companion, that can be interpreted as a vortex/dust trap at the $m=1$\ Lindblad resonance of the planet. Accretion on the candidate planet is not detected from the VLT-MUSE archival data. 
  This may be due to insufficient contrast, an observational geometry that is unfavorable for viewing the planet’s surface, or it could indicate that we are merely observing irregularities within the disk. Finally, we identify a microjet extending from the star perpendicular to the disk in these data. }


\titlerunning{WRAY15-1880: chasing the PDS 70 twin }
\authorrunning{E. Rigliaco et al.}

   \maketitle
%

\section{Introduction}

Planets form in protostellar disks, from which they accrete the material for their formation. The observation of (giant) planets that are still accreting material within their natal environment is crucial to constrain models for their formation. However, despite the large number of exoplanets observed so far, there are just a few cases where forming planets are directly observed in their natal environment. The first such objects discovered were PDS 70, which hosts two giant planets \citep{Keppler2018, Keppler2019, Haffert2019}, and HD 169142 \citep{Gratton2019, Hammond2023, Law2023ApJ}. Very recently, three additional cases have been found: 2MASS J16120668-3010270 \citep{Ginski2025}; TYC 5709-354-1: \citep{Close2025, vanCapelleveen2025}; and AB Aur: \citep{Currie2025}. Progress is slow for several reasons. The process of planet formation and accretion happens from a circumplanetary disk (CPD) which is itself embedded in a larger circumstellar and protoplanetary disk, making direct planet detection difficult. Moreover, most planets form within the closest tens of au from their host stars, making it challenging to observe them at the distances of the nearest star-forming regions. The disk may hide planets in the very early phase of formation, especially in systems seen at high inclination. In addition, the final clearing of the disk is fast, making the observation of systems in this phase quite improbable. Given the complex phenomenology expected for extrasolar planets, the addition of new cases is of paramount importance.

WRAY 15-1880 (aka RX J1842.9-3532) is a solar-type classical T Tau star in the CrA-North subregion of the CrA-complex at a distance of 151~pc \citep{Gaia2023}. \citet{Rigliaco2025} estimated a mass of 1.24~$M_{\odot}$ and an approximate age of about 3-6 Myr. The star shows a clear infrared excess that can be attributed to a pre-transitional circumstellar disk. \citet{Hughes2010} combined resolved sparse aperture masking observations and the spectral energy distribution to infer the presence of an optically thin disk within 5~au of the star and of a narrow ring of optically thick material at $\sim$0.01–0.2 au. Their model suggests little to no evidence for shadowing from the inner on the outer disk. They estimated the disk mass to be 0.01~$M_{\odot}$ and measured the inclination ($i$) to be 54~degree with a position angle ($PA$) of 32~degree. The dusty outer disk extending from 30 to 45 au was imaged by the Atacama Large Millimetre/submillimetre Array (ALMA: \citealt{Francis2020}) with a moderately long baseline. More recently, WRAY 15-1880 is among the targets of the exoALMA survey, observed with high resolution (0.074$^{\prime\prime}$), and high signal-to-noise ratio \citep{Teague2025}. Analysis of these data \citep{Curone2025} yields $PA$=26.35$^{+0.06}_{-0.06}$ degree and $i$=39.32$^{+0.03}_{-0.04}$ degree.  However, \citet{Hilder2025} found that uncertainties in stellar mass and inclination increase by an order of magnitude when a more realistic noise model is considered. The residuals of the best axis-symmetric fit of the exoALMA data show a possible spiral arm within the inner gap of the disk (\citealt{Curone2025}). The spiral arm may be caused by a giant planet possibly visible in high-contrast imaging data, if massive enough. In addition, WRAY 15-1880 was observed within the Strategic Explorations of Exoplanets and Disks with Subaru (SEEDS, \citealt{Uyama2017}) and using sparse aperture masking at the Keck telescope \citep{Willson2016}, without detection of companions.

WRAY 15-1880 was a target of the Spectro-Polarimetric High-contrast Exoplanet REsearch (SPHERE) DESTINYS Large Program (PI Ginski) that used high-contrast near-infrared polarimetric imaging revealing the disk. This imaging mode is not optimized for the detection of faint companions at very short separation. Although these data have been recently published in the comprehensive census of NIR planet-forming disks \citep{Garufi26}, we present here a detailed analysis of the polarimetric data for WRAY 15-1880. In addition, we examined two sets of observations obtained with different instruments and available from the archives: {\it i)} unpublished high-contrast imaging data acquired with SPHERE and {\it ii)} integral field spectroscopy data, assisted by adaptive optics, with the Multi Unit Spectroscopic Explorer (MUSE) at the Very Large Telescope (VLT). Moreover, we analyzed the ALMA data acquired in 2016 and used in the study by \cite{Francis2020}. We compared these data with the results obtained by the exoALMA collaboration \citep{Teague2025}. Combining all these high-quality datasets (summarized in Table~\ref{tab:obs_log}) we found several interesting features of this object that were not previously discussed. In Sect.~\ref{sect:obs} we present the data, in Sect.~\ref{sect:results} we show the results, and in Sect.~\ref{sect:discussion} the discussion. The conclusions are given in Sect.~\ref{sect:conclusions}.

\section{Observations and data analysis}
\label{sect:obs}
\subsection{SPHERE polarimetric data}

WRAY15-1880 was observed as part of the DESTINYS ESO Large Program (program 1104.C-0415 -- PI Ginski) in high-contrast imaging and polarimetry with SPHERE \citep{Beuzit2019} at the ESO-VLT telescope on May 16, 2021 (epoch 2021.37). The dual-band IRDIS \citep{Dohlen2008} subsystem was used. Observations were performed in the polarimetric imaging mode of IRDIS \citep{deBoer2020, vanHolstein2020} with broad-band H filters. This mode is optimized for disk visualization. The instrument was in pupil-tracking mode, allowing the field of view to rotate. The central star was placed behind the N\_ALC\_YJH\_S apodized Lyot coronagraph (radius 92.5 mas). A total of 104 images with an individual exposure time of 32 s were recorded. The images were split into 26 polarimetric cycles (with four half-wave-plate orientations per cycle). Conditions were good, with a median seeing of 0.60$^{\prime\prime}$ and an atmosphere coherence time of 5.3~ms. 

The IRDIS polarimetric data were reduced with the IRDAP \citep{vanHolstein2020} pipeline using the default settings. The final products included $Q_\phi$ polarized scattered light images of the disk after application of polarization differential imaging \citep{Kuhn2001}, as well as total intensity images after application of angular differential imaging \citep{Marois2006}. 
The image was recently published by \citet{Garufi26} as part of the NIR census of planet forming disks observed in near-IR high-contrast imaging. 

\subsection{SPHERE high contrast imaging data}

We also used archival SPHERE observations of WRAY 15-1880 obtained in the standard IRDIFS mode \citep{Beuzit2019}, with the integral field spectrograph (IFS: \citealt{Claudi2008}) observing in the $Y-J$ band and the dual-band imager IRDIS \citep{Dohlen2008} observing in the narrow-band $H2/H3$ filters. This mode is optimized for the search for close companions, possibly in the inner gap of the disk. 
The observations were acquired in pupil-stabilized mode to allow angular differential imaging, again with the N\_ALC\_YJH\_S apodized Lyot coronagraph. 
The star was observed twice with SPHERE, in 2016 (program 097.C-0591(A) -- PI Schmidt) and in 2018 (program 0101.C-0686(A) -- PI Schmidt). Data were not published. The second observation was obtained in poor weather conditions, yielding a contrast that is about 0.9 mag worse than the 2016 one and will not be additionally considered hereafter. The first observation was acquired at epoch 2016.38 and has the following details: nDIT$\times$DIT=36$\times$64~s, seeing=1.00$^{\prime\prime}$, field rotation=43.1 degree, average Strehl Ratio=0.43.
Standard data reduction was carried out at the High Contrast Data Centre\footnote{\url{https://hc-dc.cnrs.fr/}}, \citealt{Delorme2017, Galicher2018, Maire2016}; reduced data were kindly provided by Philippe Delorme.  
For the IRDIS data we used differential imaging obtained using the Principal Component Analysis (PCA) method after subtraction of the first five components. For IFS, we performed standard data analysis procedures developed within Osservatorio Astronomico di Padova \citep{Mesa2015} that use simultaneous angular and spectral differential imaging with the principal component method (ASDI-PCA) with 25 modes subtracted. Experiments with artificial planets showed that both IFS and IRDIS data yielded a contrast of 10.7 mag at 0.5 arcsec. Due to the information provided by the spectral dependence of speckles, the IFS contrast is better than the IRDIS one at shorter separations. At wider separations, IRDIS provides a higher contrast because of the lower impact of thermal background and read out noise.

\subsection{ALMA 2016 data}

We considered long baseline (beam of $0.16^{\prime\prime}\times 0.12^{\prime\prime}$) ALMA data that were obtained on August 26, 2016 (epoch 2016.65) and previously analyzed by \citet{Francis2020}. We considered the continuum data acquired in band 7 (centered at 870~$\mu$m) that provides emission by cold dust. We retrieved reduced data as provided by the ALMA archive. The final image was created using the {\it tclean} task in version 5.5.0 of the Common Astronomy Software Applications (CASA) package.

\subsection{MUSE data}

We retrieved optical integral field spectroscopic archive data for WRAY 15-1880 acquired on 28 Aug 2022 (epoch 2022.66) with MUSE \citep{Bacon2010} at the ESO VLT telescope (Program 109.23A6.011 -- PI Caceres). Data were not published. MUSE was used in high-resolution mode coupled with the Adaptive Optics Facility \citep{Stroebele2012, Arsenault2008}.  We retrieved reduced and calibrated data using the standard MUSE pipeline \citep{Weilbacher2020} from the ESO Science Portal\footnote{ \url{http://archive.eso.org/scienceportal}}. The data set consists of a total of about 2 hours integration time acquired with quite good atmospheric conditions (median seeing of 0.66$^{\prime\prime}$ and atmospheric coherence time of 3.8~ms). The datacube covers a total field of 11.36$^{\prime\prime}$ (diameter) with a scale of 0.0252~$^{\prime\prime}$/pixel; the spectral range is from 4749.62 to 9349.62 \AA\ with a spectral step of 1.25 \AA. After frame selection, the best reduction considers a total integration time of 3096 s. The observation was obtained in a field-stabilized mode. In this case, high-contrast imaging is possible in selected bands corresponding to emission lines by subtracting the point spread function obtained from adjacent reference wavelength bands. This observation is then optimized for the visualization of emission by gas in regions close to the star (e.g., outflows or accretion flows).

We checked that the scale and orientation of the reduced MUSE datacube are correct using the position of the background star Gaia DR3 6733646398078098688, which was projected at a separation of 3.887 arcsec and PA=75.65 degree with respect to WRAY~15-1880 at the epoch of this observation.

A schematic summary of all data used in this paper, the relative epochs of observations, and the investigated features is shown in Appendix~\ref{app:observations}, Table~\ref{tab:obs_log}.

\section{Results}
\label{sect:results}
\subsection{Stellar age}
\label{subsect:age}

The dynamical mass of the star was determined with high precision by \citet{Longarini2025} using the kinematics of the disk at a value of $M_*=1.042\pm 0.011$ \MSun. Strictly speaking, this is the mass within the gas disk seen by ALMA; it then includes not only the mass of the star but also the mass of the inner disk whose presence is inferred from the spectral energy distribution and of any possible companion that might be responsible for the gap observed in the disk. However, if we assume that this dynamical mass is indeed the mass of the star, we may constrain its age by matching the observed magnitude of the star corrected for the absorption (both interstellar and circumstellar) with predictions by isochrones. To be more conservative, we adopt as error for the dynamical mass the one suggested by \citet{Hilder2025}, who argued that the error might increase by an order of magnitude when a more realistic noise model is considered ($M_*=1.042\pm 0.11$ \MSun).
\citet{Rigliaco2025} obtained a value of $A_G=0.90\pm 0.17$ mag for absorption by comparing the observed ($B_p-R_p$) GAIA color with the spectral type of the star (the error bar is mainly due to uncertainties in the spectral type). Given the GAIA parallax, this implies absolute magnitudes of $M_G=4.925\pm 0.17$ and $M_J=3.29\pm 0.07$. Using the isochrones of \citet{Baraffe2015} we obtained the ages of $2.7\pm 1.0$ Myr from $M_G$, and $2.9\pm 0.7$ Myr from $M_J$. We averaged these values and concluded for an  age of $2.8\pm 0.7$ Myr for WRAY~15-1880. 

\subsection{A candidate companion close to WRAY 15-1880}
\label{subsection:candidate_companion}

\begin{figure*}[!ht]
    \centering
    \includegraphics[width=0.8\linewidth]{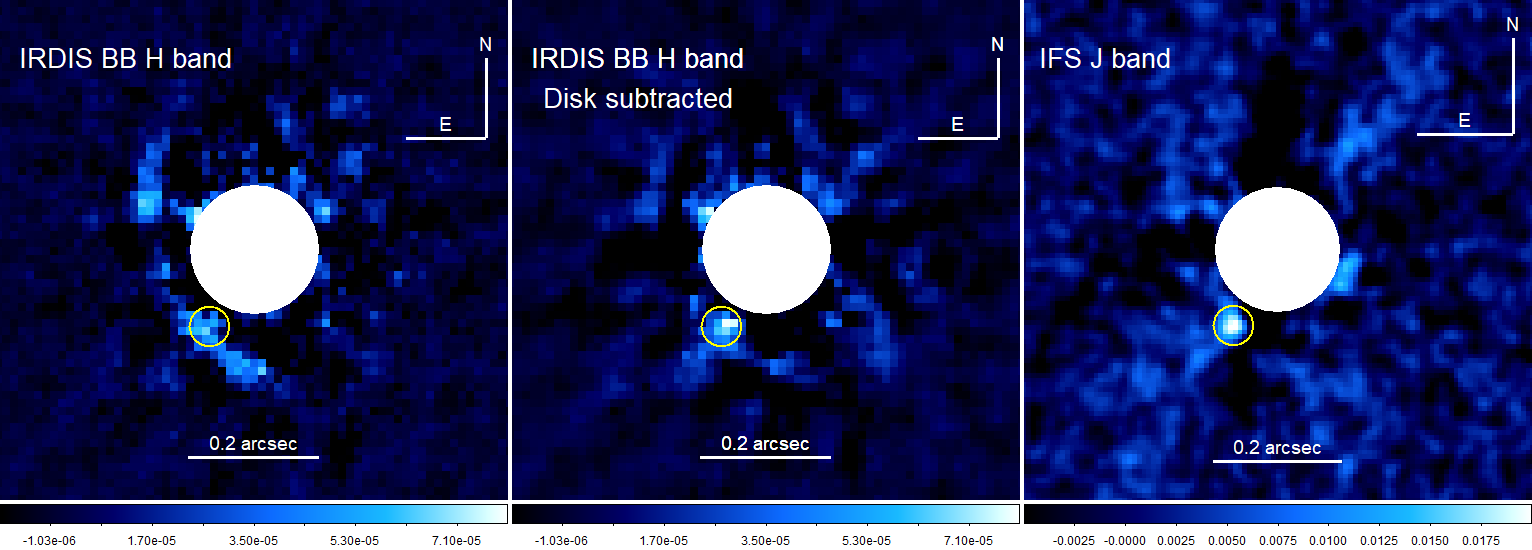}
    \caption{IRDIS and IFS contrast images of the source. Left panel: IRDIS broad band H-band contrast image. Central panel: the same, after subtraction of the disk signal. Right panel: IFS J-band contrast image. In both panels, the region within 97 mas (behind the coronagraphic mask) is masked. The yellow circle marks the possible candidate at separation of 140.8 mas and PA=150.3 degree, detected at a SNR=6.5 on the IFS image. }
    \label{fig:ifs_snr_map}
\end{figure*}

Figure~\ref{fig:ifs_snr_map} shows the IRDIS and IFS contrast images of the region around WRAY 15-1880. In the right panel, we identified a possible candidate companion with a low signal-to-noise ratio (SNR=6.5) using the IFS data set. Given the relatively large field rotation during acquisition of this data set, subtraction of the first 25 modes optimizes contrast in the range of separation from 0.1$^{\prime\prime}$ to 0.2$^{\prime\prime}$ from the star. However, the candidate is also clearly visible in the intensity images obtained after subtracting the first 10 or 50 modes (see Fig.~\ref{fig:modes} in Appendix~\ref{app:candidate_modes_spectrum}). 
The image of this source has a Full Width at Half Maximum (FWHM) of $25\pm5$~mas in right ascension and $35\pm5$~mas in declination\footnote{Errors in the FWHM are obtained from \cite{Landman1982}}, consistent within the uncertainties with the values expected for a point source at the observed wavelength ($\sim 30$~mas). The probability of having such a bright background object so close to the star is $9.4\times 10^{-6}$ using the Besan\c{c}on galactic model \citep{Czekaj2014}. Using the negative planet method, we found that if confirmed as a real physical companion, this signal should correspond to a Jupiter-like planet with a contrast of 9.03$\pm$0.54~mag in the J-band, located at an apparent separation of 140.8$\pm$4.4~mas and $PA=150.3\pm 0.7$~deg. 

The left panel of Fig.~\ref{fig:ifs_snr_map} shows the IRDIS broad H-band image. The continuous ring-like feature that will be discussed in Sect.~\ref{sect:disk} is clearly visible, and on top of that there is some signal at the location of the candidate companion (slightly closer to the star than the disk). To show this signal more clearly, we show the image obtained after subtraction of the disk signal by assuming that the disk is symmetric with respect to the semi-minor axis (central panel). The position of this possible signal (separation of 137~mas and $PA=151$~deg) is very similar to the one observed with IFS. We estimate a SNR=3.3 and a contrast of about $10.26\pm 0.34$ mag for the detected signal, showing that if the signal observed in the IFS data is due to a real companion, it likely has a blue color.
 
We obtained photometry and astrometry of this candidate companion using the negative planet method, using monochromatic images obtained after subtraction of the first 2-6 principal component analysis (PCA) modes and selecting only the $J$-band images, where SNR is highest. We found a $J$-band magnitude of $12.32\pm 0.54$. We used as error bar the standard deviation of the results obtained with different number of modes. The object was not clearly detected in the $H$-band, but the observation is compatible with an $H$-band magnitude of $12.86\pm 0.34$ simply from the SNR estimated above. Given the large errors, the $J-H$ color is compatible with a T3-type object but we cannot exclude a strong contribution by reflected starlight. The orbital parameters for this candidate companion, assuming that the orbit of the planet is circular and coplanar to the disk plane, are semi-major axis of 25.7 au and period of about 127 yr. The results are summarized in Table~\ref{tab:parameters}.  

\begin{table}[htb]
    \caption{Stellar and candidate companion properties.}
    \centering
    \begin{tabular}{lccc}
\hline
\hline
Parameter & Value & Unit & Source \\
\hline
\multicolumn{4}{c}{Star parameters}\\	
\hline
Parallax	&6.62	&mas	&Gaia	\\
$A_G$		&0.90$\pm$0.17	&mag	&R25\\					
$M_G$		&4.925$\pm$0.17	&mag	&R25\\					
$M_J$		&3.29$\pm$0.07	&mag	&R25\\
$M_H$       &2.60$\pm$0.07	&mag	&R25\\
Mass 	    &$1.042\pm0.11$ &\MSun	&L25\\	
Age		    &$2.8\pm 0.7$	&Myr	&This paper\\
\hline
\multicolumn{4}{c}{Companion parameters}\\
\hline
Separation  &$140.8\pm 4.4$  &mas	&This paper \\
PA          &$150.3\pm 0.7$  &degree &This paper \\
Semi-major axis&25.7&au     &This paper \\
Period      &127.5   &yr     &This paper \\
Contrast $J$	&$9.03\pm 0.54$	&mag	&This paper \\
$M_J$		&$12.32\pm 0.54$	&mag	&This paper \\
$M_H$		&$12.86\pm 0.34$	&mag	&This paper \\
Mass		& 0.3--7.6 &\MJup	&This paper \\
Spectral type& T3   &       &This paper \\
\hline
    \end{tabular}
    \label{tab:parameters}
    \tablefoot{The mass and spectral type of the candidate companion are obtained under the assumption that we are imaging the companion photosphere. R25 refers to \citealt{Rigliaco2025}, L25 refers to \citealt{Longarini2025}.}
\end{table}

\subsubsection{Spectrum of the candidate companion} 

\begin{figure}[!h]
    \centering
    \includegraphics[width=0.49\textwidth]{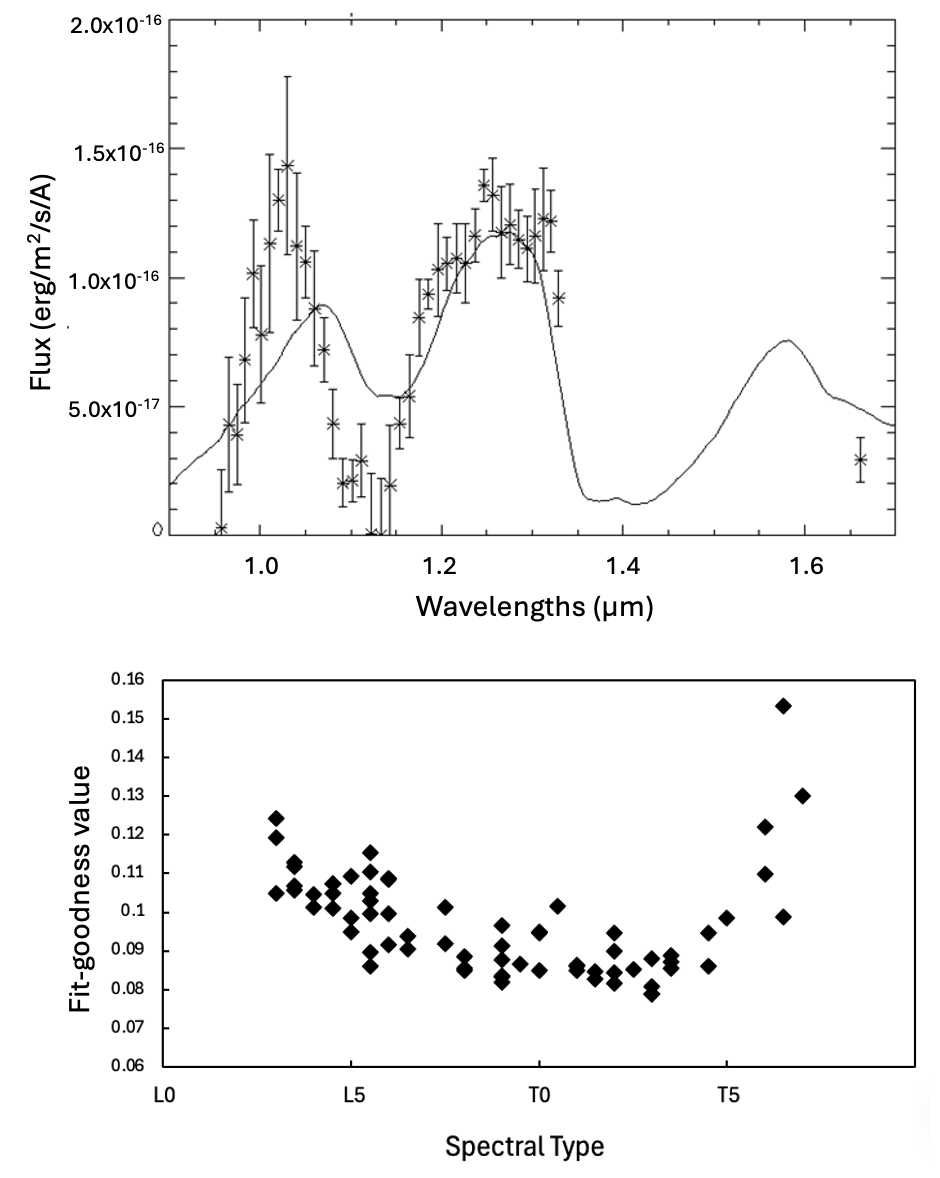}
    \caption{Top: comparison between the spectrum of the candidate companion to WRAY 15-1880 (asterisks with error bars) and the  best fitting template spectrum (SDSS1206+28, spectral type T3, from the library of \citealt{Leggett2005}). Bottom: fit-goodness value versus Spectral Type plot.}
    \label{fig:spectrum}
\end{figure}

We extracted the spectrum of the close companion candidate using the negative planet method on monochromatic PCA images. To reduce the noise, we smoothed the spectrum to reduce scatter with a current median over five spectral points, providing a fit-goodness value. 
The contrast spectrum was transformed into flux units using a fit to the primary spectrum. The resulting spectrum is shown in the top panel of Fig.~\ref{fig:spectrum}. 

We classified the spectrum assuming that what we see is the photosphere of a low-mass object. This was done by comparing the spectrum with those contained in a library of spectra by \citet{Leggett2005}. The lowest fit-goodness value was obtained for SDSS1206+28, which is classified as a spectral type T3 (bottom panel of Fig.~\ref{fig:spectrum}). This corresponds to \teff=1200 K in the table of \citet{Pecaut2013}, which is similar to the temperature of about 1300 K expected for the $J-$ band magnitude of the candidate companion when using the COND isochrones of \citet{Baraffe1998} (described in the next section). 
Considering the low SNR, other interpretations of the spectrum might be possible, although not probable. For instance, the feature seen in high-contrast imaging may include the contribution or even totally be due to stellar reflected/scattered light by dust enshrouding the object; some azimuthal irregularity in the disk structure may, in fact, lead to detection of an apparent point source after ADI. However, the shape of the spectrum seems to rule out this possibility. In the following of the paper we will discuss other features that can provide hints on the presence and mass of the candidate companion. A description of the different methods is given in the following section and are summarized in  Table~\ref{tab:mass}. In general, all these methods agree on a mass of a few \MJup. 
 
\subsubsection{Mass of the candidate companion}

{\it Companion mass estimated from the $J-$ and $H-$band magnitude}. If we assume that what we see is the photosphere of a companion, an estimate of its mass can be obtained by comparing the absolute magnitude in the $J-$ and $H-$band with isochrones from either \citet{Baraffe1998} (COND models, clean atmospheres) or \citet{Chabrier2000} (DUST models, dusty atmospheres), the latter providing higher masses. We notice that the T-type spectrum we obtain for this candidate companion favours a clean atmosphere. The resulting mass also depends on the age we assume for the companion. If it is a planet, we expect it to be younger than the star. For instance, 2.4 Myr are required to form the core of Jupiter in a standard core accretion model, as that considered by \citet{DAngelo2021}. Based on isotopic ratios measured in meteorites, \citet{Kruijer2017} argued that Jupiter's core grew to $\sim 20$~\MEarth within $<1$ Myr, followed by a more prolonged growth to $\sim 50$~\MEarth until at least $\sim 3-4$~Myr after the formation of the Solar System. According to \citet{Alibert2018}, this may be explained if, firstly, rapid pebble accretion supplied the major part of Jupiter's core mass, and secondly, slow planetesimal accretion provided the energy required to hinder runaway gas accretion during the following 2~Myr. An offset of $\sim 2$~Myr between the formation of the star and the giant planets is plausible. Given the age of WRAY 15-1880 of $2.8\pm0.7$~Myr determined in Sect.~\ref{subsect:age}, we then consider a typical age of 1 Myr for this candidate companion. We also compute the mass estimate using the nominal age of the central star ($\sim$3 Myr), as is routinely done in the literature. 
Assuming an age of 1 Myr, in $J-$band the mass of the companion is $2.7\pm0.9$\MJup using COND isochrones \citep{Baraffe1998} and $4.8\pm1.0$\MJup using DUSTY isochrones \citep{Chabrier2000}, averaging in a mass of $3.7\pm1.0$\MJup. 
In $H-$band we find $1.7\pm0.3$\MJup using COND isochrones and $2.6\pm0.3$\MJup using DUSTY isochrones, averaging in a mass of $2.1\pm0.3$\MJup. 
Assuming an age of 3 Myr, in $J-$band the mass of the companion is $4.7\pm1.3$\MJup using COND isochrones and $7.6\pm1.3$\MJup using DUSTY isochrones, Averaging in a mass of $6.1\pm1.3$\MJup. 
In $H-$band we find $3.1\pm0.5$\MJup using COND isochrones and $4.2\pm0.6$\MJup using DUSTY isochrones, averaging in a mass of $3.6\pm0.6$\MJup. \\

{\it Companion mass estimated from the size of the Hill sphere}. 
A planet rapidly accreting material from the disk is expected to be surrounded by a thick cloud of dust approximately the size of the Hill sphere \citep{DAngelo2008, Peplinski2008, Peplinski2008a, Peplinski2008b}. This cloud is larger than the proper circumplanetary disk as considered e.g. by \citet{Ayliffe2009}. If we then assume that what we see is light reflected by dust filling the Hill sphere, the ratio between the mass of the companion $M_p$ and the mass of the star $M_*$ should be approximately given by the following equation:
\begin{equation}
(M_p/M_*) \sim 24\, (10^{-0.4\, c}/A)^{3/2},
\end{equation}
where $A$ is the fraction of stellar light incident on the Hill sphere reflected towards the observer and $c$ is the contrast in magnitudes. If we assume $A=0.1$, the mass of the companion corresponding to a contrast of 9.03 mag would be $M_p=3.0^{+3.4}_{-1.6}$\MJup (similar to that obtained in the case of a naked photosphere), while it would be about a factor of ten lower for $A=0.5$. We also notice that in the first case the Hill radius would be 17 mas, meaning that the Hill sphere would have a diameter comparable to the SPHERE resolving power, and it would be about half that for $A=0.5$. This result suggests that even if we are observing stellar light scattered by dust around a compact object filling its Hill sphere, this is likely a planet.\\

{\it Companion mass estimated from the size of the gap}. 
The candidate planet is in the disk gap. If its orbit is nearly circular and on the same plane of the outer disk, then the semi-major axis is 25.7 au, closer to the outer edge of the gap (which is between 27 and 30 au depending on the data set considered). The inner edge of the gap (the outer radius of the inner disk) is much closer to the star: $\leq 13.1\pm 1.5$ au. This might suggest that the planet is actually closer to the star than what is obtained in this way. This may occur if the orbit is eccentric (and the planet was in 2016 close to the apocentre) or the orbital plane has a lower inclination than the outer disk, or both. In all cases, a better determination of the orbit of the candidate planet would be very useful. 
However, we could still provide a rough estimate of the ratio $q$ between the mass of the planet carving the gap observed in the disk of WRAY 15-1880 and the star using the formula of \citet{Kanagawa2016}: $q=2.1\times 10^{-3}\,(W/a_p)^2 (h_p/0.05\, a_p)^{1.5} (\alpha/10^{-3})^{0.5}$. In this formula, $W=101$~mas is the gap size (the difference between the inner radius of the outer disk $R_{\rm in}$(outer)$=188$~mas and the outer radius of the inner disk $R_{\rm out}$(inner)$=87$~mas), $a_p=170$~mas is the semimajor axis of the orbit of the planet, $h_p/a_p\sim 0.1$ is the height of the disk at the position of the planet and $\alpha$ is the viscosity of the disk, which we assume to be 0.001 \citep{Francis2020}\footnote{This values also agrees with the empirical relation between accretion rate on the star and disk viscosity found by \citet{Rafikov2017}}. We get a mass ratio of $q=0.0021$ from the formula of \citet{Kanagawa2016}, which implies a mass of about 2.2~\MJup for the planet. The value is uncertain due to the assumptions of the circular orbit formula in the eccentric scenario, but we can also assume that the candidate planet is on a circular orbit, and in this case the half-width of the gap would be the distance between the inner rim of the outer disk (188~mas) and the planet (170~mas), corresponding to a gap size $W=36$~mas. The corresponding mass ratio is in this case $q=0.0003$ from the formula of \citet{Kanagawa2016}, meaning M$_p\sim$2.2~\MJup,  almost an order of magnitude lower than the previous estimate.

\begin{table*}[!ht]
\caption{Summary of the different mass estimates that can be obtained for the candidate companion using different methods. The column "Remarks" highlights the major shortcomings of each method.}
    \centering
    \begin{tabular}{|l|c|c|}
\hline
\hline
Method & Mass (\MJup) & Remarks  \\
\hline
$J$- and H-band mag &     $1.7-7.6\pm 1.5$    & Model/Age dependent; Star light contribution. \\ 
Hill sphere size    & $3.0^{+3.4}_{-1.6}$ & Photospheric contribution; Filling of Hill sphere.  \\
Gap size        &          0.3--2.2        & Uncertain calibration; disk height; disk viscosity.     \\
\hline
\hline
    \end{tabular}
\label{tab:mass}
\end{table*}

\subsection{The disk around WRAY~15-1880}
\label{sect:disk}

The analysis of the VLT-SPHERE polarimetric data and ALMA 2016 data indicate the presence of two sections of the disk: an inner section, whose presence is also indicated by the spectral energy distribution (e.g., \citealt{Francis2020}) and the clear signs of accretion on the star (e.g., \citealt{Pittman2025}), and an outer disk. Even if the inner disk is mainly hidden behind the coronagraph, there might be some evidence of its presence in the polarimetric data from light spilling out of the mask, while the outer disk is clearly detected both in SPHERE and ALMA images. In the following sections, we show the modelling of the SPHERE polarimetric images, and of the ALMA data, performed to gain insight in the morphology of the inner and outer disk around WRAY~15-1880. 

\subsubsection{Features in the polarimetric images}
\label{sect:featurespolarimetry}

\begin{figure*}[!ht]
\centering
\includegraphics[width=0.8\textwidth]{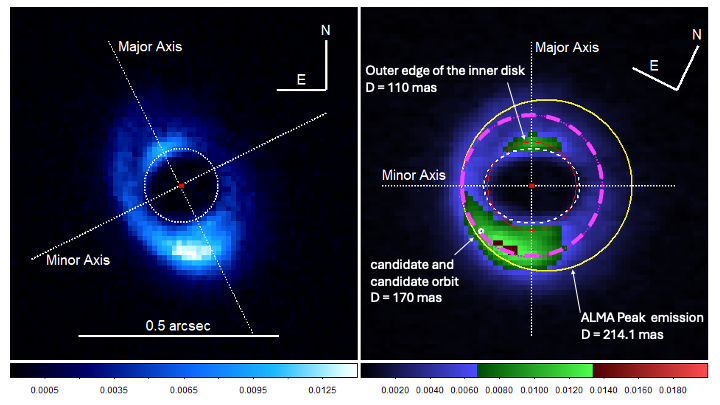}
   \caption{Left: $Q_\phi$ image of WRAY 15-1880 obtained from the SPHERE polarimetric data. The red dot at the center of the image marks the position of the star; the dashed white circle is the size of the coronagraph used in this observation. The dotted straight lines mark the major and minor axes, as obtained by \citet{Curone2025}. Right: similar image, but projected on the disk plane ($i=39.22$~deg) and rotated to have the major axis vertical and the minor axis horizontal. On this plane, the coronagraphic mask is an ellipse represented by the white dashed line. The yellow circle has a radius of 214.1 mas that is the peak of the disk as determined by the ALMA data (see Section 3.3.2). The center of this circle has been offset by 37 mas along the minor axis, to fit the outer contour of the observed disk East of the star, that is the closest side of the disk. The red solid circle (radius of 110 mas) fits the feature seen north of the star. The position of the planet discussed in Section~\ref{subsection:candidate_companion} at the epoch of the SPHERE polarimetric observation is marked with a small white circle, and its orbit is shown as magenta circle.}
    \label{fig:qphi_image}
\end{figure*}

\begin{figure*}[!ht]
    \centering
    \includegraphics[width=0.7\textwidth]{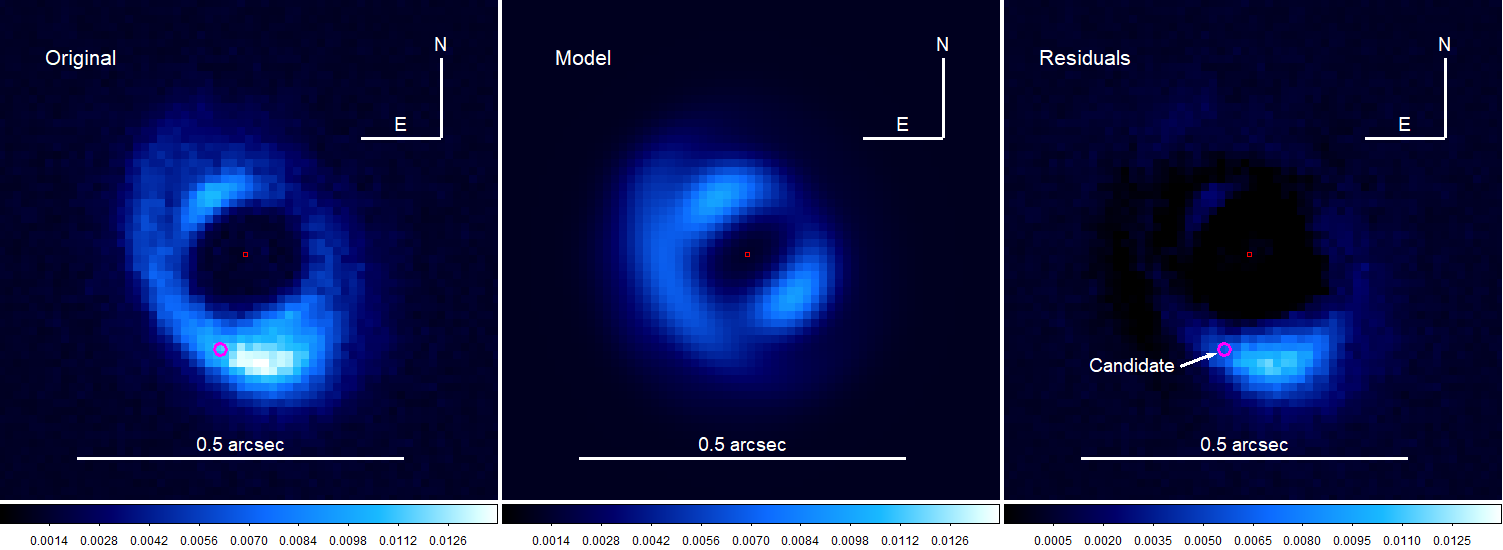}
    \caption{Left panel: $Q_\phi$ image of WRAY 15-1880 obtained from the SPHERE polarimetric data. The red dot at the center of the image marks the position of the star; the position of the candidate planet at the epoch of the SPHERE polarimetric observation is marked with a small magenta circle. Middle panel: the scattering model considered for the disk (see text). Right panel: residuals}
    \label{fig:scattering_model}
\end{figure*}

Figure~\ref{fig:qphi_image} in the left panel shows the $Q_\phi$ image of WRAY 15-1880 obtained from the SPHERE polarimetric data. The red dot in the center of the image marks the position of the star; the dashed white circle is the size of the coronagraph used in this observation (92.5~mas). The dotted lines mark the major and minor axes, as obtained by \citet{Curone2025}. 
The right panel shows the $Q_\phi$ image projected in the disk plane, using the inclination angle of 39.22~degree computed by \citet{Curone2025}, and rotated so that the major axis is vertical and the minor axis horizontal. In this plane, the coronagraphic mask is an ellipse represented by the white dashed line. The yellow circle has a radius of 214.1~mas (32.3~au), which is the peak height of the disk as determined by the ALMA data (see Section~\ref{subsection:alma2016data}). 
The light-solid red circle (radius 110 mas) fits the feature seen both North and South of the star close to the coronagraphic edge. This might represent the outer edge of the inner disk, which is mostly hidden behind the coronagraphic mask and then only barely visible in this image. The dotted magenta circle marks a possible circular orbit in the disk plane (radius of 25.7 au, which is 170 mas at the distance of WRAY 15-1880) of the candidate planet discussed in Section~\ref{subsection:candidate_companion}. The position of this planet at the epoch of SPHERE polarimetric observation is marked with a small white circle. We notice that with an outer radius $<110$ mas, the inner disk is contained within the $m=2$ inner Lindblad resonance of this candidate planet, expected at 107 mas for a Keplerian disk. We also notice that this disk is not evident in the ALMA data, suggesting that the inner disk is poor in large grains.

Moreover, in Figure~\ref{fig:qphi_image}, right panel, the yellow circle matches the position of the outer ring, but the center of this circle has been offset by $dx=37$~mas west along the minor axis to fit the outer contour of the observed disk East of the star, which is the side closest to us. This offset may be because light is scattered by grains at the surface of the disk, rather than settled in the disk midplane. Given the thickness of the disk, the surface is elevated above the midplane (see \citealt{Ginski2016}). The elevation $h$ of the disk's surface over the plane is then expected to be given by $h=dx/\sin{i}$, where $i$ is the inclination. If we adopt the inclination determined by \citet{Curone2025} ($i=39.22$ degree), we obtain $h=55$ mas. This implies that the form factor of the disk is $h/R\sim 0.26$. A more precise value will be determined in the next Section. It is not surprising that this is larger than the value of $h/R=0.114\pm 0.019$ obtained from the ALMA data (see Section~\ref{subsection:alma2016data}) because the SPHERE polarimetric image gives the disk thickness for micrometre grains and the ALMA data for grains of millimetre size. The latter are expected to be more settled towards the disk plane than the others.

To determine more accurate values for the disk parameters, we built a simple scattering model from the surface of the disk. 
We assume a bow-tie shape of the disk, where the height of the disk is linearly proportional to the distance, based on a Henyey-Greenstein scattering function \citep{Henyey1941} characterized by the $g$--factor. 
The modeled disk has an aperture (disk shape factor) $h/r$ and is cut between the minimum and maximum radii $R_{min}$ and  $R_{max}$. We neglect the contribution by the inner wall of the outer disk because we assume that it is hidden by the inner disk, and the contribution of the inner disk because it is hidden behind the coronagraphic mask. A grid of 62,800 points, in cylindrical coordinates equi-spaced in angle (step of 1 degree) and radius (step of 12.25~mas), is drawn on the disk surface. By construction, all points on the grid have the same illumination. The relative brightness of each point of the grid is then given by the scattering phase function, which takes into account the angle ($\psi$) between the star and the observer (computed using $i=39.22$ degree and $PA=26.35$ degree from \citealt{Curone2025}) as seen from each point in the grid using the Henyey-Greenstein function. The image in the disk plane is then projected onto the sky plane. The brightness in scattered light in the final image is the sum of the contribution of all points in the original grid that fall within each individual detector pixel as seen from the observer. Finally, the image obtained this way is multiplied by the transmission of the coronagraphic mask as determined by A. Boccaletti (private communication), convolved with the instrumental profile (assumed to be a Gaussian with a standard deviation $\sigma$), and normalized to the total intensity of the observed image. The free parameters in the model are $R_{min}$, $R_{max}$, $g$, $h/r$, and $\sigma$. We extracted a random set of 10,000 values for these parameters with a uniform wide distribution and produced model scattering images for each of them. The best solution is the one that minimizes the standard deviation of the difference between the model and the data in a specified region of the image (see below). Uncertainties on the final parameters used in the fit are obtained from the standard deviation of the parameters for those solutions where the standard deviation is less than 10\% more than the value obtained with the best solution. 

As a first step, we focused on the outer disk. We optimized the parameters by minimizing the scatter in the residuals between observations and the model in the North-East sector of the disk ($-26.35<\phi<63.65$) that is not affected by the bright region South of the star. With this procedure, we obtain the following best values for the outer disk parameters: $R_{\rm min}=188\pm 14$ mas (28.4$\pm$2.1~au), $R_{\rm max}=218\pm 17$ mas (32.9$\pm$2.6~au), $h/R=0.15\pm 0.07$, and $g=0.93\pm 0.05$. As expected, the value of $R_{\rm max}$ is similar to the value of the peak of the disk seen by ALMA as determined in Section~\ref{subsection:alma2016data} ($R_0=214.1\pm 2.4$ mas). The high value of $g$ indicates the dominance of forward scattering.

The inner disk, if detected, is at the very edge of the inner working angle of the coronagraph. Nevertheless, we attempted to model it with the same procedure described above. We fixed the position angle by assuming that the inner disk is perpendicular to the gaseous jet ($PA_{\rm in}=27.63$\ degree, discussed in Sect.~\ref{subsection:accr_microjet}) and left the inclination free because it is possible that the inner and outer disks are misaligned. We arbitrarily fixed the inner radius ($R_{\rm min}=24.5$ mas) and the disk shape factor ($h/R=0.1$) because the presence of the coronagraph does not allow us to determine their values. We optimized the parameters for the inner disk by minimizing the scatter in the residuals in the northern half of the disk (not affected by the bright region south of the star) and with a distance from the star between the edge of the coronagraph (92.5 mas) and a radius of 154 mas. The values we obtained have only qualitative meaning, with an inclination of the inner disk of the order of $i\sim 28\pm 3.0$ degrees, $R_{\rm max}\leq 87$~mas (13.1~au) and $g=0.40\pm 0.29$ (the last parameter is essentially unconstrained). According to this fit, the inner disk is seen at a lower inclination than the outer disk; the outer radius of the inner disk is comparable to the radius of the coronagraphic mask explaining why it is only barely visible close to the major axis, and the scattering phase function of the inner disk is more isotropic than in the outer disk. The results are summarized in Table~\ref{tab:disk_parameters}. 

The sum of the models for the inner and outer disks is shown in the middle panel of Figure \ref{fig:scattering_model}. The residuals between the $Q_\phi$ image and the sum of the models displayed in the right panel of this figure show an obvious peak centered at $PA=176.1$ degree, close to the position of the candidate planet discussed in Sect.~\ref{subsection:candidate_companion}, expected for the epoch of the polarimetric observation ($PA=166.6$ degree). 
 In the same figure we notice that the distribution of the residuals is not symmetrical. A possible interpretation of this asymmetry is discussed in Appendix~\ref{app:spirals}. 

\subsubsection{Features in the ALMA data}
\label{subsection:alma2016data} 

\begin{figure}[htb!]
    \centering
    \includegraphics[width=0.45\textwidth]{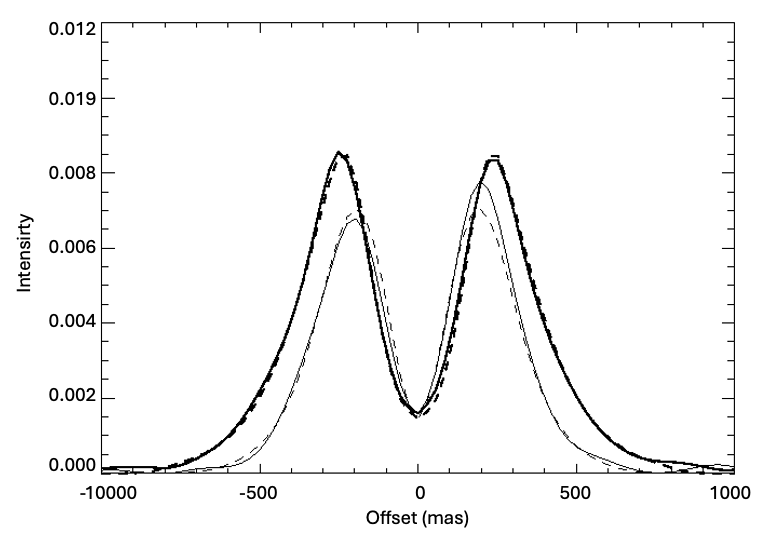}
    \caption{Comparison between the disk profile for WRAY 15-1880 observed with ALMA (solid line) and the geometric model considered in this paper (dashed line). Thick lines refer to the profile along the major axis; thinner lines refer to the profile along the minor axis.}
    \label{fig:disk_profile}
\end{figure}

\begin{figure*}[htb!]
    \centering
    \includegraphics[width=0.9\textwidth]{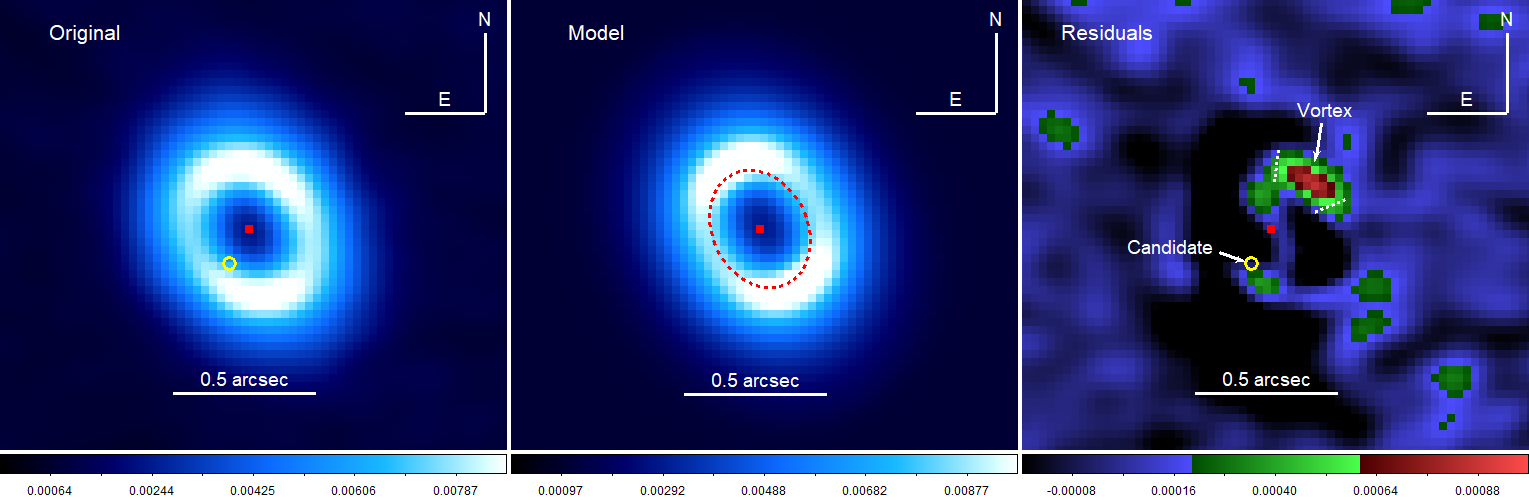}
    \caption{Left panel: the disk of for WRAY 15-1880 as observed by ALMA in the 870 $\mu$m continuum. In all panels, the red square marks the position of the star and the open yellow circle the expected position of the candidate planet observed with SPHERE at the epoch of the observations, assuming a circular Keplerian motion on the disk plane. The dashed red ellipse show the position of the disk peak. Central panel: the geometric model for the dust distribution. Right panel: residuals between observation and model. The positions of the vortex and of the candidate companion are also marked. Note that the color scale (that is shown on bottom of each panel) was multiplied by 8 in this last image to show faint details. 
    }
    \label{fig:disk_residuals}
\end{figure*}

We analyzed the cold dust emission observed with ALMA in 2016 with a simple geometrical model: we assumed that the emission is proportional to the density $\rho$ of the dust in the disk and that this is distributed symmetrically around its axis according to the following law:
\begin{equation} \label{eq1}
\begin{split}
\rho (r,h) = & \rho_0\,\exp{-[(h/r)/h_0]}\, \exp{-[(R_0-r)/\delta_{\rm in}]}\\
&~~~~~~~~~~~~~~~~~~~~~~~~~~~~~~~~~~~~{\rm if}~ r<R_0  \\
\rho (r,h) = & \rho_0\,\exp{-[(h/r)/h_0]}\, \exp{-[(r-R_0)/\delta_{\rm out}]} \\
&~~~~~~~~~~~~~~~~~~~~~~~~~~~~~~~~~~~~{\rm if}~ r>R_0 \\
\end{split}
\end{equation}
where $r$ and $h$ are the radial and vertical coordinates (in mas), $\rho_0$ is the peak density in the disk plane at a radius $R_0$, and $\delta_{\rm in}$ are the decaying radii on the inner and outer sides of the peak and $h_0$ with height (in mas). We also considered that the disk is seen at an angle $i$ and has a position angle on the sky of $PA$, so the signal in the ALMA image depends on the integral along the line of sight for each point of the disk. We adopted the inclination and position angles considered in the disk model of WRAY 15-1880 by \citet{Curone2025} ($i=39.22$ degree and $PA=26.35$ degree). 
We also assumed that the disk may be optically thick. In our model, the optical thickness $\tau$ is related to the surface luminosity by a factor $w=\exp{[-|(z/(r\, \tau)| )]}$, where $z$ is the vertical coordinate (with respect the disk mid-plane and along the line of sight) and $r$ is the radial coordinate (on the disk plane). Finally, the model profile was convolved with a bi-dimensional Gaussian reproducing the ALMA beam as recorded in the header of the file. The parameters considered in this model were optimized to reduce the standard deviation of the residuals between observation and the model, after appropriate normalization of the total intensity. We find $R_0=214.1\pm 2.4$ mas (32.3$\pm$0.4~au), $\delta_{\rm in}=49.1\pm 1.3$ mas, $\delta_{\rm out}=115.4\pm 2.9$ mas, $h_0=0.164\pm 0.031$ (corresponding to an aspect ratio of $0.114\pm 0.019$ at half-peak intensity), and $\tau=0.166\pm 0.003$. The maximum value of the correction factor due to the optical depth is  $w=1.181\pm 0.004$\footnote{With these values, the half-intensities of the disk are $R_{\rm in}=180.1\pm 1.8$ mas and $R_{\rm out}=294.1\pm 3.1$ mas}. The root mean square of the residuals is 0.025 of the maximum intensity of the disk.

Figure~\ref{fig:disk_profile} compares the observed profile of the apparent luminosity of the disk along the major and minor axes with the prediction of the simple model considered above. Although the fit is generally good, there are clear residuals. The right panel of Figure~\ref{fig:disk_residuals} shows the distributions of these residuals in the sky. They are not distributed symmetrically with respect to the star, showing that an axis-symmetric model cannot exactly reproduce the data. These residuals show an arc-shaped excess of emission North-West of the star with a peak SNR=4.9. This is projected within the disk and is not coincident with the minor axis. Hence, it cannot be identified with an outflow from the star. An emission excess at a similar (though not identical) position is also visible in the 2022 data analyzed by \citet{Curone2025} supporting its robustness. In the 2016 observation the arc-shaped residual covers the position angle $293<PA<352$ degrees (roughly opposite to the position of the candidate planet at the same epoch) and is centered at an apparent separation of 206 mas. If in the plane of the disk, this corresponds to a feature at a separation of about 259 mas (39.1 au) that is about 1.52 times the distance of the candidate planet proposed in the previous subsection from the star and then close to the $m=1$\ outer Lindblad resonance, which is expected at 1.587 times the planet orbit in a Keplerian disk. 
This feature might be interpreted as a dust trap generated by a large-scale vortex at this resonance \citep{Pinilla2012, vanderMarel2013, Owen2017}. To support this interpretation, we analyzed the ALMA data acquired in 2022 with the same model described above, measuring the rotation that occurred to this feature between 2016 and 2022 using a cross-correlation of the signal in the two epochs, in the region of the vortex employing the angle as independent variable. If the feature were in Keplerian rotation, at the observed distance, we should expect a rotation of $\sim$8.8 degrees. If the feature were instead dynamically linked to the candidate companion and solidly corotating, we expect a rotation of $\sim$16.6~degree. The observations show a rotating angle of 15.9$\pm$2.6~degrees, in agreement with the corotating dynamically linked scenario. This result supports the interpretation of this feature as being located at the $m=1$\ outer Lindblad resonance.

\subsection{Accretion on the star and microjet detection from MUSE data}
\label{subsection:accr_microjet}

\begin{figure*}[htb]
    \centering
    \includegraphics[width=0.9\textwidth]{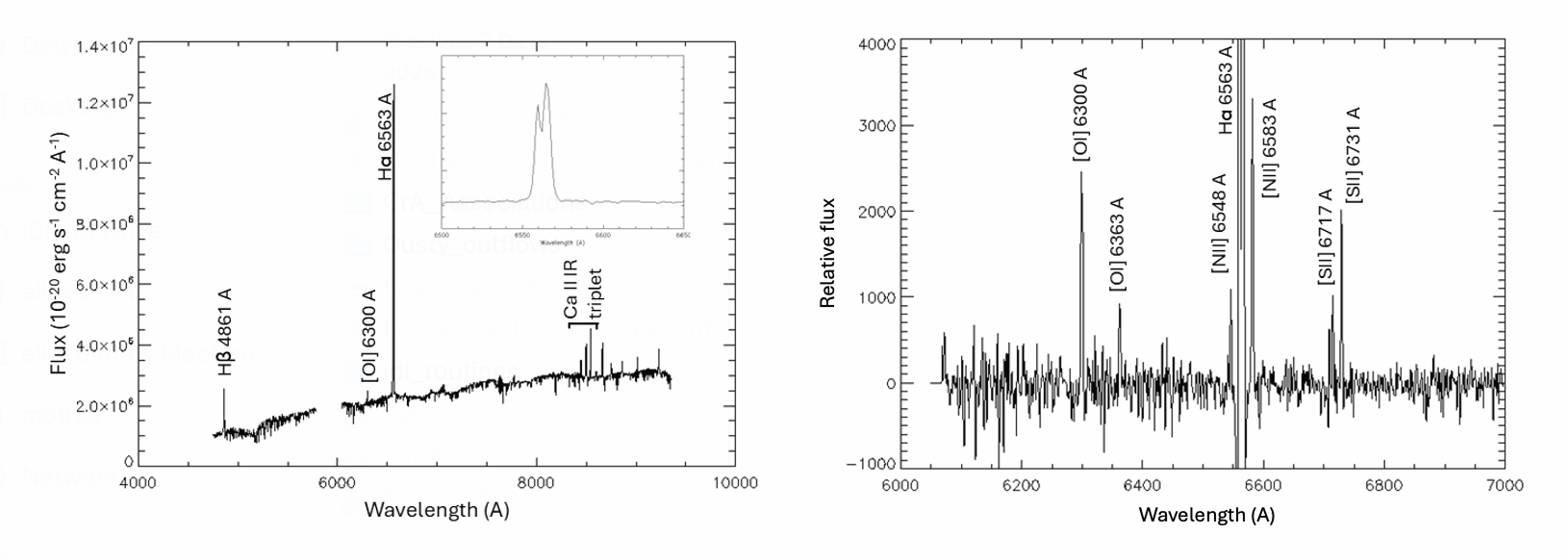}
    \caption{Left: MUSE spectrum of WRAY 15-1880, with blow-up around the H$\alpha$ line. Right: region between 6000--7000~\AA\, after subtracting the stellar spectrum in the same region. The spectrum is extracted in a region between 150--450~mas from the central star, and constant aperture of 225 mas. The emission lines due to a microjet are highlighted and annotated. }
    \label{fig:muse_spectrum}
\end{figure*}

The left panel of Figure \ref{fig:muse_spectrum} shows the spectrum of WRAY 15-1880 extracted from the MUSE data. Some emission lines are visible (Hydrogen Balmer and Paschen lines, [O\enspace I] at 6300\AA, the near-IR Ca\enspace II triplet at 8498\AA, 8542\AA, and 8662\AA, among others). 
While the profiles of the Balmer lines show two peaks, the profile of the [O\enspace I] line at 6300\AA\ is unresolved in this low resolution spectrum. We used the H$\alpha$ and H$\beta$ lines to estimate the accretion rate on WRAY 15-1880 from the MUSE spectrum. By integrating along the line profile and subtracting the local continuum, we estimate total fluxes F$_{H\alpha}$=$1.54\times10^{-12}$~erg/s/cm$^{2}$ and F$_{H\beta}$=$3.86\times10^{-13}$~erg/s/cm$^{2}$. Given the distance from the star and assuming $A_G=0.9$ mag \citep{Rigliaco2025}, the corresponding luminosities are  $L_{H\alpha}=2.6\times10^{-3}$~\LSun and  $L_{H\beta}=8.6\times10^{-4}$~\LSun, producing the accretion luminosities of $\log{L_{\rm acc}/L_\odot}=-0.98$ and $\log{L_{\rm acc}/L_\odot}=-0.91$ using the relations of \citet{Fang2009}. If we now assume a mass of 1.042~\MSun and radius 1.75~\RSun for WRAY 15-1880 as appropriate for the star age of 2.8 Myr, and the \citet{Baraffe2015} isochrones, the accretion rates are $3.6\times10^{-9}$~\MSun/yr (that is, $\log{{\dot M}/M_\odot}=-8.44$) from H$\alpha$ and $4.3\times10^{-9}$~\MSun/yr (that is, $\log{{\dot M}/M_\odot}=-8.37$) from H$\beta$ using the relation between the luminosity and the accretion rate of \citet{Rigliaco2012}. The average is $\log{{\dot M}/M_\odot}=-8.40\pm 0.04$\footnote{We obtained a very similar value of $\log{{\dot M}/M_\odot}=-8.32\pm 0.14$ using the Ca\enspace II lines at 8542 and 8662~\AA\ and the calibration of \citet{Herczeg2008}.}. For comparison, \citet{Pascucci2007} measured $\log{{\dot M}/M_\odot}=-9.0$ (assuming a distance of 140~pc and an interstellar absorption $A_V=1.03$ mag); \citet{Manara2014} measured $\log{{\dot M}/M_\odot}=-8.8$ (assuming a distance of 130 pc and an interstellar absorption $A_V=0.40$ mag); \citet{Pascucci2020} obtained $\log{{\dot M}/M_\odot}=-8.51$ (assuming a distance of 153 pc and an interstellar absorption $A_V=0.6$ mag). All these results were obtained from an estimate of the accretion luminosity based on a combination of X-ray (derived from ROSAT observation), optical, and infrared (Spitzer) data, taking into account distance, interstellar absorption, and stellar radius and mass. Differences in accretion rates might be explained by variability; however, we also notice that we assume a higher interstellar absorption than adopted by \citet{Pascucci2020} and a larger distance and a higher interstellar absorption than considered by \citet{Manara2014}.

\begin{figure}[htb]
    \centering
    \includegraphics[width=0.5\textwidth]{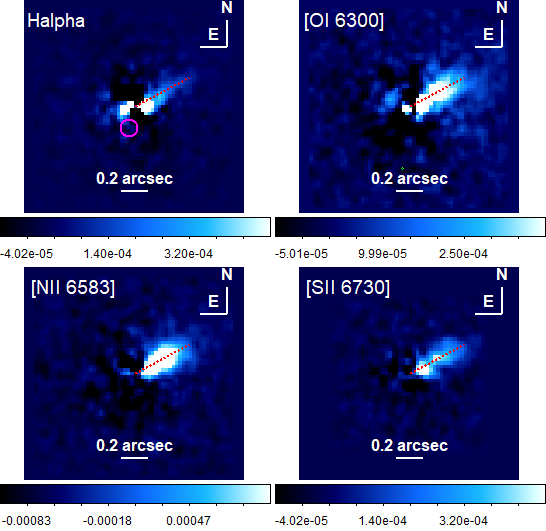}
    \caption{Images obtained from MUSE data for WRAY 15-1880 in H$\alpha$ (upper left panel), [O\enspace I] 6300 \AA\ (upper right panel), [N\enspace II] 6583 \AA\ (lower left panel), and [S\enspace II] 6730 \AA\ (lower right panel) lines. Images are in units of contrast with respect to the local continuum. In all cases the stellar PSF obtained from adjacent continuum wavelengths and the first three principal components have been subtracted. The solid (dashed) red arrow indicates the position of the front side of the polar microjet from the star with PA=$297.63\pm 0.37$ degrees. The red circle in the upper left panel mark the position expected for the candidate planet possibly present in the SPHERE data at the epoch of the MUSE data}
    \label{fig:muse_images}
\end{figure}

Figure \ref{fig:muse_images} shows the images in H$\alpha$, [O\enspace I] at 6300 \AA\, [N\enspace II] at 6583 \AA, and [S\enspace II] at 6730 \AA\ obtained from the MUSE datacube. In all cases, the point spread functions obtained from adjacent continuum spectral regions and the first three components of a PCA on the residuals were subtracted to show regions with flux excess. These emission maps clearly show the contribution from a polar microjet North-West of the star (directed toward the observer and projected on top of the distant side of the disk) that is detectable up to about 0.44~arcsec from the star. 

This microjet is also likely responsible for the broad blueshifted component of the [O\enspace I] 6300\AA\ observed by \citet{Banzatti2019} in the high-resolution spectrum of WRAY 15-1880. The portion of the spectrum of this microjet between 6000--7000\AA\ (after subtracting the stellar spectrum in the same region) is shown in the right panel of Figure \ref{fig:muse_spectrum}. It shows prominent H$\alpha$ and forbidden lines of [O\enspace I], [N\enspace II], and [S\enspace II], indicating a low-density region. Using averages over the entire jet, we measured ratios of 3.03 between the intensity of the [N\enspace II] at 6583 \AA\ and 6548 \AA\ (close to the expected theoretical values \citealt{Storey2000}) and of 0.51 between the [S\enspace II] lines at 6717 \AA\ and 6731 \AA. The presence of the [N\enspace II] lines indicates a density below the critical value of $6.6\times10^4$cm$^3$ \citep{Osterbrock2006} and the low ratio obtained for the [S\enspace II] lines indicates a higher density than $10^4$cm$^3$ \citep{Kewley2019}. The average density within the microjet is therefore restricted by these line ratios to be in the range $1-6\times10^4$cm$^3$. This range is consistent with the intensity ratio of 0.713 between the [Fe\enspace II] lines at 7155\AA\ and 8617\AA, which indicates a density of $4.0\times10^4$cm$^3$ (see \citealt{Podio2006}). This density value is within the range observed for microjets in protostars \citep{Lavalley-Fouquet2000, Podio2006}.

We determined the position angle and the projected semi-aperture (1$\sigma$) of the microjet from the images in each of the lines by means of a cross-correlation with a mask describing the projected microjet. The weighted average of the position angles ($297.63\pm 0.37$ degrees) is consistent within the error bars with the position angle of the disk determined by \citet{Hughes2010} (PA = 32 degrees), but given the small error bar, it is significantly different from that expected from the more accurate determination obtained from the ALMA data (PA =$26.35\pm 0.06$ degree, \citealt{Curone2025}). This suggests some misalignment between the inner and outer disks. However, the offset between the $PA$\ of the microjet and the normal to the disk in the case of WRAY 15-1880 ($1.3\pm 0.4$ degrees) is within the range typically observed in classical T Tau stars (e.g. \citealt{Flores-Rivera2023}). 

Finally, we used the observed wavelengths of the emission lines related to the microjet, as well as the wavelengths of a number of photospheric absorption lines, to measure the relative radial velocity of the microjet with respect to the star. This is $-110\pm6$\kms at a typical projected separation of about 0.2$^{\prime\prime}$ from the star\footnote{For comparison, the blue-shifted component of the [O\enspace I]~6300\AA\ line found by \citet{Banzatti2019} has a radial velocity of -121~\kms with respect to the star. This possibly reflects the velocity of material in a portion of the microjet that is closer to the star than what is considered here.}. If we assume that the disk inclination is $i=39.22$ degree (as given by the analysis of the ALMA data, \citealt{Curone2025}) the microjet semi-aperture and the corresponding internal error is $4.56\pm 0.08$~degree, and the microjet velocity relative to the star is $142\pm8$\kms. These values for the microjet semi-aperture and velocity are consistent with expectation of an unconfined supersonic flow for the typical temperature of $T\sim10^4$ K, within the range usually considered for microjets from young stars. 

\section{Discussion}
\label{sect:discussion}

\subsection{Comparison with exoALMA disk}

\begin{table}[htb]
    \caption{Comparison between disk parameters obtained from fitting of the SPHERE polarimetric images and from ALMA 2016 and 2022 data. }
    \centering
    \small
    \begin{tabular}{l|cc|c}
\hline
\hline
Parameter	    & Polarimetry   & ALMA 2016 & ALMA 2022 \\
 &  \multicolumn{2}{c}{(This paper)} & \citep{Curone2025}\\	
\hline
\multicolumn{4}{c}{Inner disk} \\			
\hline
$PA$ (deg)     & {\it 27.63} & & \\	
$i$	  (deg)         & $28\pm 3$ & & \\		
$R_{\rm out}$ (mas)    & $<87$   & & \\		
$R_{\rm out}$ (au)    & $<13.1$    & & \\		
\hline
\multicolumn{4}{c}{Outer disk} \\			
\hline
$PA$ (deg) & {\it 26.35} & {\it 26.35}   & $26.35\pm 0.06$\\
$i$	  (deg) & {\it 39.22} & {\it 39.22}   & $39.22\pm 0.04$\\
$R_{\rm in}$ (mas)	& $188\pm 14$ & $180.1\pm 1.8$ & $200\pm 0.9$  \\
$R_{\rm in}$ (au)	& $28.4\pm 2.1$ & $27.2\pm 0.3$ & $30.2\pm 0.2$  \\
$R_{\rm peak}$ (mas)	& $218\pm 17$ & $214.1\pm 2.4$ & 	237        \\
$R_{\rm peak}$ (au)	& $32.9\pm 2.6$ & $32.3\pm 0.4$ & 	35.8        \\
$R_{\rm out}$ (mas)    &             & $294.1\pm 3.1$ & 	293        \\
$R_{\rm out}$ (au)    &             & $44.4\pm 0.4$ & 	44.3       \\
\hline
    \end{tabular}
    \label{tab:disk_parameters}
    \tablefoot{Values in italics without error bars are assumed.}
\end{table}

\begin{figure*}[htb]
    \centering
    \includegraphics[width=0.75\textwidth]{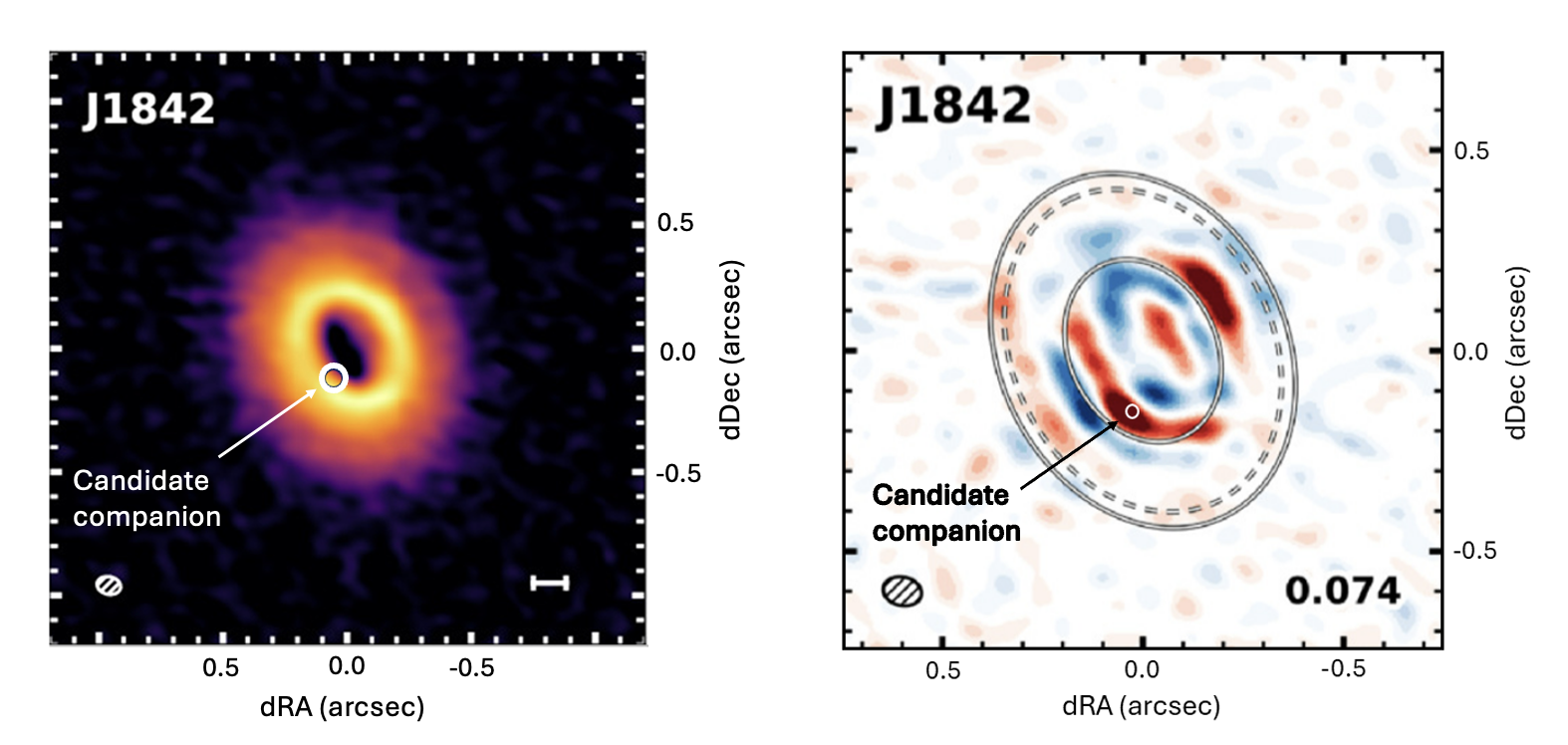}
    \caption{Readapted from \citealt{Curone2025}. Left: ALMA disk continuum image of WRAY 15-1880 at 0.9~mm (331.6 GHz) from exoALMA survey \citep{Curone2025}.  Right: residual plot generated by subtracting the frank model (sampled at the same uv-points of the observations) from the data and then imaged with CLEAN \citep{Curone2025}. The color scale represents the values of residual SNR in units of the observed noise (rms). In both panels the white circle represents the expected location of the candidate companion at the epoch of the ALMA observation, assuming a circular Keplerian motion on the disk plane.}
    \label{fig:alma}
\end{figure*}

ALMA data for WRAY 15-1880 were also obtained within the exoALMA program. \citet{Curone2025} computed an axisymmetric model of the disk and derived the main parameters. Table \ref{tab:disk_parameters} compares the disk parameters obtained from the analysis of the SPHERE polarimetric images with those obtained by the ALMA data (both the exoALMA and the 2016 data analyzed in this paper).

\citet{Curone2025} found significant residuals from the axisymmetric model suggesting a spiral arm on the East of the star (although other interpretations are possible) and a peak West of the star at PA$\sim 309$ degrees. We may compare the map of the residuals obtained by \citealt{Curone2025} with the position of the planet expected at the epoch of the ALMA data. These last data were collected on April 20, 2022 (\citealt{Teague2025}, epoch 2022.30), corresponding to 5.92 yr after the SPHERE observation. Assuming a  counter-clockwise \citep{Izquierdo2025} circular Keplerian motion around the star on the disk plane between the two epochs, the planet should be at a separation of 151.6 mas at $PA=167.9$ degree at this epoch. The result of this comparison is shown on the right panel of Figure~\ref{fig:alma}. The position of the candidate companion matches very well with the maximum of the residuals East of the star, suggesting that these residuals are indeed due to material related to the candidate planet. In addition, the position of the West peak reproduces that of the peak seen in the residuals of the 2016 ALMA image; this supports the real presence of this feature.

Finally, there is no evidence for a molecular outflow from ALMA data for WRAY 15-1880 discussed in the exoALMA survey; the lack of a molecular outflow even in the presence of microjets is a quite common feature among classical T Tau stars \citep{Shang2023}. 

\subsection{Properties of the candidate companion}

Table~\ref{tab:mass} summarizes various determinations of the mass of the candidate companion of WRAY 15-1880 that we discussed in this paper. Each of them has significant methodological uncertainties, but, on the other hand, the values obtained for the candidate companion are not too far from each other, consistently suggesting a mass of a few \MJup for the candidate companion. A large set of distinct observables is explained by a planet similar to or slightly larger than Jupiter around WRAY 15-1880.  Although this is not a clear proof that this planet exists, it makes it a good candidate. We might then ask ourselves if we should expect an H$\alpha$ emission from this candidate companion. 

\begin{figure}[!ht]
    \centering
    \includegraphics[width=0.9\linewidth]{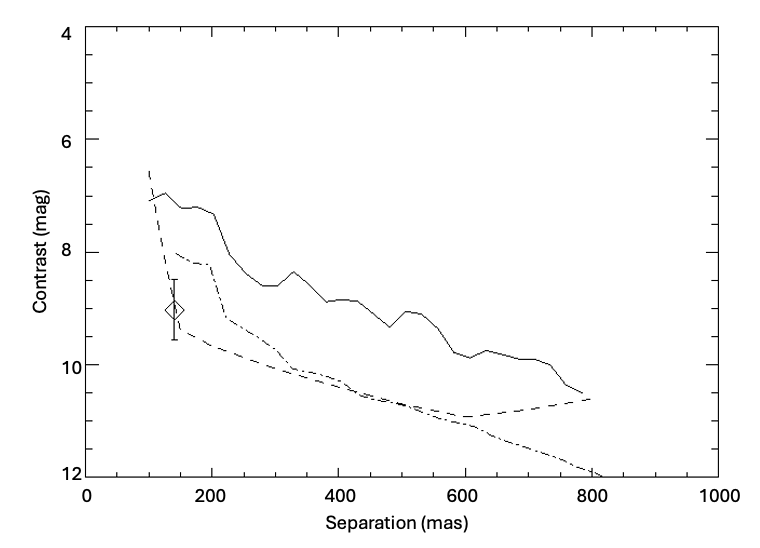}
    \caption{5$\sigma$ contrast in H$\alpha$ obtained from the MUSE image (solid line). For comparison we also show the contrast in $J-$band (dashed line) and in $H-$band (dot-dashed line) obtained with SPHERE in 2016. The diamond symbol with error bar marks the contrast and separation of the candidate planet observed with SPHERE in the $J-$band.}
    \label{fig:contrast}
\end{figure}

An accreting companion such as the candidate described in Section 3.2 may be bright in H$\alpha$ and then detectable in MUSE data \citep{Haffert2019}.  In Fig.~\ref{fig:contrast} we plot the 5$\sigma$ contrast curve that we obtain from the H$\alpha$ image. The contrast takes into account self-subtraction inherent to the PCA algorithm. This was computed considering the attenuation of the signal from simulated planets (only emitting in H$\alpha$) injected into the images. This is compared with the similar curve in the $J-$ and $H-$ bands obtained from the SPHERE data; we also marked in this figure the separation and contrast of the possible candidate planet found in the last data. From this figure we found that if the contrast in H$\alpha$ is similar to the contrast of the candidate planet possibly detected in the SPHERE $J-$band image, it would not be detectable in the MUSE data. We set an upper limit of about 7.5 mag to the contrast in $H_\alpha$ at the expected position for the planet, corresponding to L${_{H\alpha}}<2.6\times 10^{-6}$\LSun for the H$\alpha$ luminosity of the planet. The corresponding accretion rate needed to produce this value of L${_{H\alpha}}$ can be obtained using the formulas of \citet{Aoyama2018}, for a 3~\MJup planet with radius 1.82~\RJup (the value of a 1~Myr old planet of this mass using the isochrones of \citealt{Baraffe1998}). The upper limit on the H$\alpha$ luminosity implies an upper limit of $9.7\times10^{-7}$ \MJup/year to the accretion rate on this putative planet. This value is 58 times higher than the accretion rate estimated by \citet{Aoyama2019} for the planets of PDS 70. For reference, the stellar accretion rate we found for WRAY 15-1880 ($\log{{\dot M}/M_\odot}=-8.40\pm 0.04$) is about 30 times higher than that for PDS 70 (variable in the range of $\log{{\dot M}/M_\odot}$ from -9.7 to -10.2 : \citealt{Thanathibodee2020}) in agreement with its younger age. 

\subsection{Distribution of material in the planet orbit}

\begin{figure}[htb]
    \centering
    \includegraphics[width=0.9\linewidth]{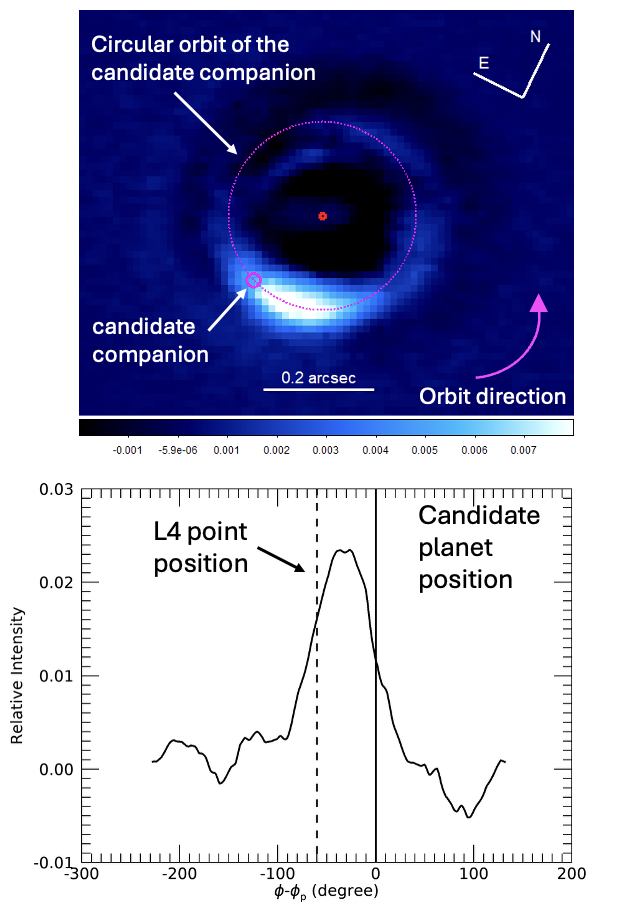}
    \caption{Top panel: residuals of the $Q_\phi$ image after subtraction of the model of scattering by the disk of WRAY 15-1880, projected on the disk plane. The small red circle represents the position of the star; the small magenta circle the position of the candidate planet discovered on the SPHERE high-contrast image; the large magenta dashed circle is a circular orbit passing through its position. Bottom panel: intensity along the circular orbit from the image on the top panel as a function of the angle $\phi$. This is the angle with respect to the star in this plane computed counterclockwise starting from the major axis North of the star. The solid vertical line represents the planet position; the dashed vertical line represents the position of the L4 Lagrangian point}
    \label{fig:type_iii}
\end{figure}

The top panel of Figure~\ref{fig:type_iii} shows the projected residuals after subtracting the scattering model by the disk surface (Sect.~\ref{sect:featurespolarimetry}) from the $Q_\phi$ image projected on the disk plane. If the candidate planet will be confirmed, this indicates the presence of an excess of material co-orbiting it on the leading side of the orbit. The bottom panel of this figure shows the relative intensity along the orbit as a function of the angle $\phi$. This is the angle with respect to the star in this plane computed counterclockwise starting from the major axis North of the star; $\phi_p$ is the value of $\phi$ for the planet. The excess extends over the region from the planet position up to about $\phi_p-90$ degrees, slightly further than the L4 Lagrangian point. 

The excess of co-rotating material on the leading side of the orbit of a forming planet corresponds to the prediction by the hydrodynamical model of planets undergoing gas accretion considered by \citet{DAngelo2008} in the case of an initial, relatively fast, planetary radial motion. This mechanism works for rather small planets ($M_P<2$ \MJup, \citealt{Pierens2016}), which is compatible with current data for the candidate companion (e.g., see Table~\ref{tab:mass}), agreeing with a very young age for this candidate planet.

\section{Conclusions}
\label{sect:conclusions}

WRAY 15-1880 (also known as RXJ1842.9-3532) is a solar-type star that belongs to the CrA-complex and is surrounded by a prominent circumstellar and protoplanetary disk. The analysis of new and archival data has led to important new discoveries on this object.  
The availability of an accurate dynamical mass of $1.042\pm 0.011$ \MSun for this star from the analysis of the disk kinematics by \citet{Longarini2025}, for which we adopted the error of 0.11\MSun as suggested by \citet{Hilder2025}, allows us to determine an age of $2.8\pm 0.7$~Myr by comparisons with the isochrones of \citet{Baraffe2015}. This makes this star very interesting because it is in the late phases of the disk evolution when giant planets form. In fact, we expect giant planets to form from 1 to 3 Myr after the star \citep{Kruijer2017} and then continue to accrete for a few million years after \citep{Alibert2018}. In addition, \citet{Longarini2025} estimate a mass of $0.078\pm 0.014$ \MSun for the (outer) disk. This gives a potential reservoir for further growth of any planet in the system. WRAY 15-1880 is therefore a laboratory for observing the formation of giant planets. Finally, \citet{Rigliaco2025} attributed the relative motion of CrA-Main with respect to CrA-North to the effects of a supernova that exploded about 3.72 Myr ago within the latter association at a distance of a few pc from WRAY 15-1880. WRAY 15-1880 might be a very intriguing analog of the very young Solar System because the isotopic abundances of several elements in carbonaceous chondrite meteorites (which likely reflects that in the proto-solar nebula) suggest the explosion of a supernova in the neighborhood of the Sun cradle \citep{Looney2006, Banerjee2016, Sieverding2020}.

We combined high-contrast and polarimetric data in the near-IR for WRAY 15-1880 acquired with SPHERE and AO-assisted integral field spectroscopy obtained with MUSE, with sub-mm observations acquired with ALMA in two epochs (2016 and 2022). The polarimetric data indicate the presence of two sections of the disk, an inner one (whose presence is also indicated by the spectral energy distribution and the clear signs of accretion on the star), and an outer one also observed by ALMA. Moreover, the high-contrast IFS $J-$band data reveal a candidate companion possibly orbiting WRAY 15-1880. The contrast is $9.03\pm 0.54$ mag in the $J-$band at an apparent separation of $140.8\pm 4.4$ mas and $PA=150.3\pm 0.7$ degree. If we are observing the photosphere of a companion, this object is a giant planet with a mass between $1.7 - 7.6 \pm 1.5$ \MJup. The spectrum of the source indicates a T3 spectral type (corresponding to a temperature of about 1200 K), again matching expectations for a planet with the observed absolute magnitude of this candidate object. 
However, we cannot completely exclude that the object imaged in high-contrast imaging results from a combination of emission by a photosphere with starlight reflected/scattered by dust enshrouding the planet, even if this possibility is unprobable because the spectrum of the candidate companion is not compatible with scattered starlight.
Alternatively, the feature may be an artifact due to irregularities in the disk structure, but also in this case, the detection at low-SNR of the candidate companion in J-band, where the disk is not visible tends to makes this interpretation unlikely. 

The accretion rate onto the star measured from the Hydrogen lines in the MUSE spectrum is $\log{{\dot M}/M_\odot}=-8.40\pm 0.04$. The MUSE data reveal the presence of a high velocity microjet ($164\pm 22$\kms) at $PA=299.4\pm0.7$ degrees with an aperture of $\sim9.1\pm0.1$ degrees. Line strength ratios indicate a typical density in the range $(1-6)\times 10^{-4}$ cm$^{-3}$ within the jet. 

The MUSE data do not show a signal that could be interpreted as due to accretion on the candidate planet, perhaps because the observation is not deep enough, and only a planet with a very high accretion rate ($\sim 10^{-6}$\MJup/yr) would be visible in this data. 
We notice that other planets within disk gaps lack evidence for accretion (e.g. HD169142b and AB Aur b). This point is discussed in \citet{Close2025}. They argue that possibly only planets that are seen at an intermediate inclination are detectable in H$\alpha$ because this makes it easier to observe the shock on the surface of the planet. In this respect, the WRAY 15-1880 disk is also seen at an intermediate inclination, so this argument would not be valid here. However, the planet might be embedded by a thick envelope that absorbs light emitted at H$\alpha$.

If the feature detected in the high-contrast images acquired with SPHERE is a real physical companion on a circular orbit around the primary, on the plane of the disk, its orbital semi-major axis is of 25.7~au and the period is about 127 yr. 

The ALMA data (both in the 2016 and 2022 data sets) indicate the presence of bright residuals over an extended arc or segment of a spiral. In the 2016 observation, it covers the position angle $293<PA<352$ degrees and is centered at an apparent separation of 206~mas. If in the plane of the disk, this corresponds to a feature at a separation of about 259 mas from the star, that is about 1.52 times the distance of the candidate planet from the star, close to its m=1 outer Lindblad resonance. We may interpret it as a dust trap generated by a large-scale vortex at this resonance \citep{Pinilla2012, vanderMarel2013, Owen2017}. This is confirmed by the rotation of this structure between the 2016 and 2022 epochs. We obtain a value of 15.9$\pm$2.6 degrees, consistent with the rotation of 16.6 degrees expected if the vortex is rotating solidly with the candidate planet as expected in Lindblad resonance; rather, it is inconsistent with the expected Keplerian rotation at the vortex position (8.9 degrees). The vortex is then a density wave and not a self-gravitating cloud.

We conclude that we find an intriguing candidate for a very young Jupiter-like planet around this star interacting with the disk around the star. This would be the most primitive forming planetary systems discovered so far. However, accretion of gas on this object is not detected in the MUSE data perhaps because they are not deep enough for this purpose. If the candidate giant planet were confirmed by further observations, WRAY 15-1880 would add to the short list of stars hosting a disk and planets imaged simultaneously, the second one after AB Aur where a microjet is also observed \citep{Rodriguez2014}. These last features can be related to the very young age of WRAY 15-1880, which appears indeed to be lower than the age of all other stars known to host accreting planets.

\begin{acknowledgements}
The paper is based on observations made at the European Southern Observatory, under programs 1104.C-0415 (PI Ginski), 097.C-0591 (PI Schmidt), 101.C-0686 (PI Schmidt), and 109.23A6.011 (PI Caceres). Moreover, it makes use of the following ALMA data: ADS/JAO.ALMA\#2015.1.01083.S. ALMA is a partnership of ESO (representing its member states), NSF (USA) and NINS (Japan), together with NRC (Canada), NSTC and ASIAA (Taiwan), and KASI (Republic of Korea), in cooperation with the Republic of Chile. The Joint ALMA Observatory is operated by ESO, AUI/NRAO and NAOJ.
This work has used the High Contrast Data Centre, jointly operated by OSUG/IPAG (Grenoble), PYTHEAS/LAM/CeSAM (Marseille), OCA/Lagrange (Nice), Observatoire de Paris/LESIA (Paris), and Observatoire de Lyon/CRAL, and supported by a grant from Labex OSUG\@2020 (Investissements d’avenir – ANR10 LABX56). 
We thank Christian Ginski for leading the DESTINYS Large Program that provided the high-contrast imaging polarimetric data analyzed in this manuscript. 
E.R. acknowledges support from the Large Grant INAF 2022 “YSOs Outflows, disks and Accretion (YODA): towards a global framework for the evolution of planet forming systems” and from PRIN-MUR 2022 20228JPA3A “The path to star and planet formation in the JWST era (PATH).” The work was also partially funded with an INAF "Mini-Grant" RF 2022. 
S.D. and E.R. acknowledge support from the “Programma di Ricerca Fondamentale INAF 2023” of the Italian National Institute of Astrophysics (INAF Large Grant 2023 “NextSTEPS”). 

\end{acknowledgements}

\bibliographystyle{aa} 
\bibliography{biblio} 

\begin{appendix}

\section{Observations }
\label{app:observations}

Table~\ref{tab:obs_log} summarizes all the new and archival data used in this study, with the corresponding observation periods and the features revealed and investigated in this manuscript. The observations are also described in Section~\ref{sect:obs}. 

\begin{table*}[!h]
    \caption{Observation log of the new and archival data employed in this paper.}
    \centering
    \begin{tabular}{|c|c|c|}
\hline
\hline
 Instrument	& Epoch & Features revealed or investigated \\
\hline
SPHERE HCI  & 2016.38 & Candidate companion \\
ALMA        & 2016.65 & Outer disk; disk discontinuities  \\
SPHERE HCI/Polarimetry  & 2021.37 & Inner and outer disk \\ 
ALMA        & 2022.30   & Outer disk; disk discontinuities. \\ 
MUSE        & 2022.66   & Accreting candidate companion \\
\hline
\hline
    \end{tabular}
    \label{tab:obs_log}
\end{table*}

\section{Detection of the candidate companion}
\label{app:candidate_modes_spectrum}

\begin{figure}[!ht]
    \centering    
    \includegraphics[width=1.0\textwidth, angle=270]{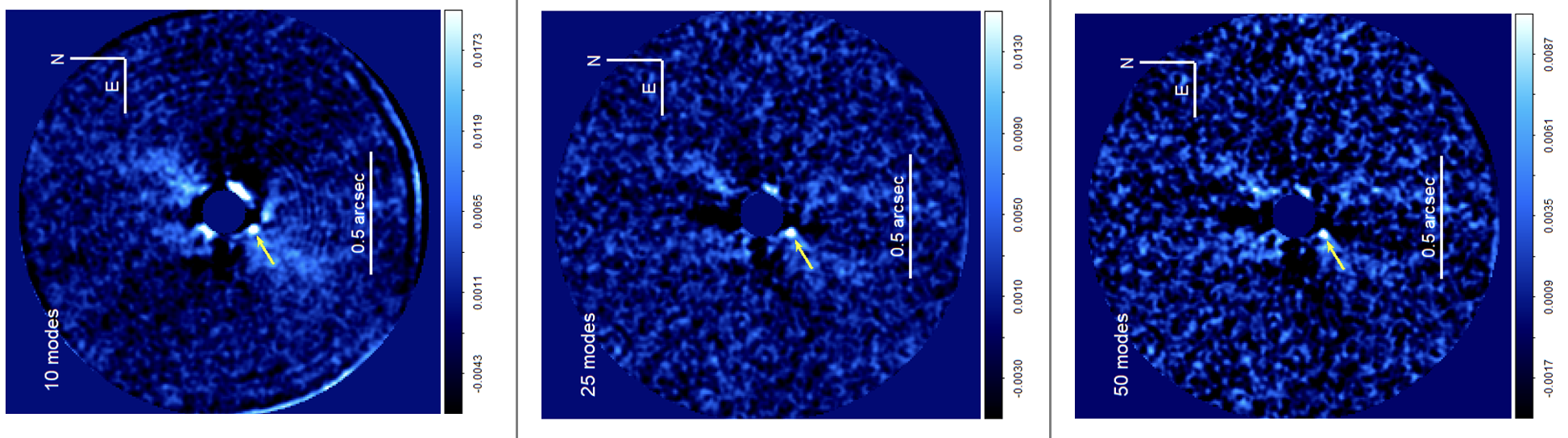}
    \caption{Intensity image in the $J-band$ of the field around WRAY 15-1880 from the 2016 SPHERE observation after application of the ASDI PCA data analysis technique \citep{Mesa2015}. Top panel: image obtained after subtraction of the first 10 modes. Middle panel: image obtained after subtraction of the first 25 modes. Bottom panel: image obtained after subtraction of the first 50 modes. North is up, East is left. The regions within 13 pixels ($\sim$97 mas, that is behind the coronagraphic mask) and beyond 115 pixels (858 mas, that is beyond the IFS field of view) from the star were masked. In all panels a yellow arrow indicates the candidate planet. }
    \label{fig:modes}
\end{figure}

The candidate companion detected in the SNR map in $J-$band at low signal-to-noise ratio (SNR=6.5) is clearly visible after subtracting the first 10, 25 and 50 modes, as shown in Figure~\ref{fig:modes}. 

\section{Possible spiral arms around WRAY15-1880}
\label{app:spirals} 

\begin{figure*}[!h]
    \centering
    \includegraphics[width=0.9\linewidth]{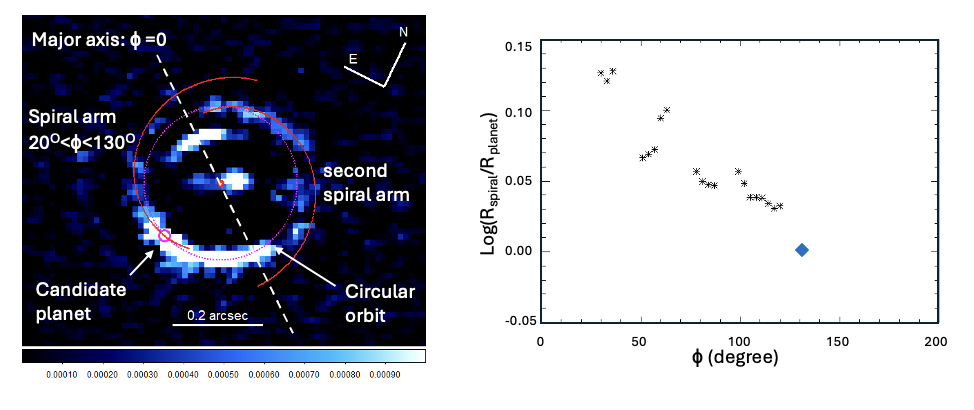}
    \caption{Upper panel: residuals after subtraction of the disk model from the $Q_\phi$ image of WRAY 15-1880 on the disk plane. A high pass filtering have been applied by subtracting the median over 7 pixels, to enhance structures. The red circle marks the star position; the magenta circle the position of the candidate planet at the epoch of this observation. The dotted magenta line is a possible circular orbit for this planet. The red solid lines mark possible spiral arms around the star. They are at 140 degree one from the other. Lower panel: peak of the possible spiral arm feature as a function of the azimuth angle $\phi$ with respect the star on the disk plane (asterisks). $\phi$ is defined from the northern position of the major axis towards East. The open diamond marks the planet position.}
    \label{fig:spiral}
\end{figure*}

A planet may generate spiral arms within a disk \citep{Zhu2015}. These may be relatively faint structures that need some manipulation of data to be clearly shown. The residual of the Q$\phi$ image of WRAY 15-1880, obtained after subtraction of the disk model described in Sect.~\ref{sect:featurespolarimetry}, show a clear asymmetry, and in the following we explore the possibility that such asymmetry may be caused by the presence of spiral arms. 
We transformed the coordinate system to image these residuals on the disk plane. To further enhance any  structure, we filtered the low frequencies on this residual image by subtracting the median over 7 pixels. The result of this manipulation is shown in the upper panel of Fig.~\ref{fig:spiral}. We find that a spiral arm can be followed in this image in the range $20<\phi<130$ degrees. We fit this structure as a logarithmic spiral. To determine the pitch angle of this spiral, we computed the position of the peak of a Gaussian fitting the radial profile of this structure at steps of 3 degrees around the star (roughly corresponding to one pixel along the spiral arm). The results of this procedure are shown in the lower panel of Figure \ref{fig:spiral}. This plot shows that a logarithmic spiral is indeed a good fit to the data. The pitch is given by the slope of the straight line that fits the data. We find that the pitch angle is $9.0\pm 0.9$ degrees in the section of the spiral arm with $20<\phi<130$ degrees. It should be noticed that this value of the pitch angle depends on the assumption that the spiral is coplanar with the disk mid-plane. If we now consider the models computed by \citet{Zhu2015}, we find that the low value of the pitch angle agrees with a perturbation in the linear regime, suggesting a mass ratio $q<0.006$ between the planet and the star, which means a mass of the planet $<$6.3 \MJup. 

We also notice that a second fainter spiral arm can also be detected in the image shown in the upper panel of Figure~\ref{fig:spiral}, with an offset of $\Delta \phi=140\pm 5$ degrees from the main arm (identified in Fig.~\ref{fig:spiral}). The offset between the two spiral arms corresponds to a mass ratio of $q\sim 0.0049\pm 0.0007$ (planet mass $\sim$5.1$\pm$0.7~\MJup) using the calibration of \citet{Fung2015}: $q/0.001 \sim (\Delta \phi/102)^5$. These values are then compatible with a giant planet with a mass of $2.5\pm 0.5$ \MJup at the location expected for the candidate planet possibly detected in the high-contrast imaging data.

\end{appendix}
\end{document}